\documentclass[11pt]{article}
\usepackage{amssymb,amsmath}
\include{epsf}
\usepackage{color}

\usepackage{graphicx}
\usepackage{amsfonts}

\newcommand{\e}{{\mathrm e}}
\newcommand{\ii}{{\mathrm i}}
\newcommand{\dd}{{\mathrm d}}
\newcommand{\eqn}[1]{(\ref{#1})}

\def\appendix#1{\addtocounter{section}{1}\setcounter{equation}{0}
\renewcommand{\thesection}{\Alph{section}}
\section*{
\thesection\protect\indent \parbox[t]{11.715cm} {#1}}
\addcontentsline{toc}{section}{Appendix\thesection\ \ \ #1} }
\newcommand{\real}{{\mathbb R}} 

\newcommand{\be}{\begin{equation}}
\newcommand{\ee}{\end{equation}}
\newcommand{\beq}{\begin{equation}}
\newcommand{\eeq}{\end{equation}}
\newcommand{\bea}{\begin{eqnarray}}
\newcommand{\eea}{\end{eqnarray}}
\def\beqa{\begin{eqnarray}}
\def\eeqa{\end{eqnarray}}
\def\nn{\nonumber}
\newcommand{\del}{\partial}

\newcommand{\eq}{\begin{equation}}
\newcommand{\eqa}{\begin{eqnarray}}
\newcommand{\en}{\end{equation}}
\newcommand{\ena}{\end{eqnarray}}

\begin{document}
\begin{titlepage}
\begin{flushright}

\baselineskip=12pt ICCUB-12-050\\
LPT-Orsay 12-09
\end{flushright}

\begin{center}

\baselineskip=24pt

{\Large\bf Gauge and {Poincar\'e}  Invariant
{Regularization}\\ and Hopf Symmetries}

\baselineskip=14pt

\vspace{1cm}

{Fedele Lizzi$^{1,2,3}$ and Patrizia Vitale$^{1,2,4}$}
\\[6mm]
$^1${\it Dipartimento di Scienze Fisiche, Universit\`{a} di Napoli
{\sl Federico II}}
\\[4mm]
$^2${\it INFN, Sezione di Napoli}\\
{\it Monte S.~Angelo, Via Cintia, 80126 Napoli, Italy}
\\[4mm]
$^3$ {\it Departament de Estructura i Constituents de la Mat\`eria,
\\Institut de Ci\'encies del Cosmos,
Universitat de Barcelona,\\
Barcelona, Catalonia, Spain}
\\[4mm]
$^4$ {\it Laboratoire de Physique Th\'eorique,\\
Universit\'e Paris Sud XI, Orsay, France}
\\{\small\tt
fedele.lizzi@na.infn.it, patrizia.vitale@na.infn.it}

\end{center}

\vskip 2 cm

\begin{abstract}
{We consider the regularization of a gauge quantum field theory
following a modification of the Pochinski proof based on the
introduction of a cutoff function. We work with a Poincar\'e
invariant deformation of the ordinary point-wise product of fields
introduced by Ardalan, Arfaei, Ghasemkhani and Sadooghi, and show
that it yields, through  a limiting procedure of the cutoff
functions, to a regularized theory, preserving all symmetries at
every stage. The new gauge symmetry yields  a new Hopf algebra with
deformed costructures, which is inequivalent to the standard one.}
\end{abstract}

\end{titlepage}

\section{Introduction}

One of the most fascinating aspects of field theory is the renormalization programme and
the role of the renormalization group. The regularization of field theory always
requires the introduction of a scale, which may be taken as a momentum scale.
In~\cite{Polchinski} Polchinski introduced  a momentum cutoff function to find an exact
renormalization group equation for a scalar theory.  The procedure is of not easy
applicability to gauge theories. This was performed later and a review of these methods
with extended references can be found in~\cite{reviews}. The extension to nonabelian
theories has remained problematic for long.  Some results have been obtained
\cite{rgnonabelian}, paying various prices like the necessity of an infinite number of
colours,  or  the loss of Poincar\'e invariance (see for example ~\cite{Gu}).
In~\cite{BoniniDAttanasioMarchesini} it is introduced a formulation for which, although
at any given scale the theory is not gauge invariant, the limiting theory is invariant.

In our case, the symmetry is maintained at all scales. The enhancement is due to the
presence of an extra hidden Hopf symmetry and the use of a deformed product. For a different approach on the
preservation of local an global symmetries  see  \cite{Morris:2005tv} and refs. therein.

Let us consider  the following action (written in momentum space)
for quantum electrodynamics for which a regularization has been
implemented by a  cutoff on the momenta:
\beqa
S_{\mathrm{QED}}&=& \int_\Lambda \frac{\dd^4 p}{(2\pi)^4}
{\Biggl\{ }
\bar{\tilde\psi}(-p)(\gamma_\mu p^\mu+m)\tilde \psi(p)
 +\; e
\bar{\tilde\psi}(-p)\int_\Lambda \frac{\dd^4q}{(2\pi)^4}
\Theta_\Lambda(p-q) \gamma^\mu\tilde A_\mu(p-q)
\tilde \psi(q) \bigr.\nn\\
&& + \;\;  \frac{1}{4} \left(\tilde F_{\mu\nu}(-p)\tilde F^{\mu\nu}(p)+\frac{1}{2\xi}
p_\mu \tilde A^\mu(-p)\,p_\nu \tilde A^\nu(p) \right) \Biggr\}.
\label{actioncutoffsharp}
\eeqa
Here by $\int_\Lambda \dd^4 p$ we mean $\int_0^\Lambda p^3\dd p$
(we are in the Euclidean case), and $\Theta$ is the characteristic
function of the four-sphere of radius $\Lambda$
\be
\Theta_\Lambda(p)=\left\{\begin{array}{cc}1\ & p^2<\Lambda\\
0\ & p^2\geq \Lambda\end{array}\right.
\ee
The slight modification with respect to usual regularizations will be clear in the next
sections. The theory described by~\eqn{actioncutoffsharp} is free of UV divergences.
 The problem with cutoff regularization is that it
destroys gauge invariance. In this paper we will show that such a regularized  action
has an hidden symmetry, which is a deformation of the usual U(1) symmetry of QED. The
new symmetry is actually a Hopf symmetry with deformed costructures, which results to be
inequivalent to the standard one. The tools we will be using have been developed for
noncommutative quantum field theory (for a review see \cite{Szabo}), although we will
only deal with commutative field theory.

Our starting point is the article~\cite{iraniani2}, where it  was
shown that a suitable commutative deformation of the  product among
fields, inspired by noncommutative quantum field theory gives a new
action, deformed by the presence of ``cutoff functions'', which
preserves gauge invariance, although in a deformed version. As long
as the cutoff functions are analytic the new action is equivalent to
the ordinary one, being related by a field redefinition. For the
action not to be just a redefinition a non-analytic cut off is
requested, which however destroys the associativity of the new
product.  We show in this paper that the desired result, a
regularized gauge invariant action, can be obtained with a limiting
procedure. Our strategy will be the following.  We start by imposing
the usual desirable space-time symmetries - translation and Lorentz
invariance - and we find the form of the deformed product compatible
with them. It is the  commutative, non-local product
introduced in~\cite{iraniani2} on the basis of~\cite{GLV09}. The new
action endowed with such a product is the deformed action considered
in~\cite{iraniani2}. We then make the cutoff function converge
towards a function which vanishes for momenta larger than a cutoff,
and hence cannot be analytic. In this limit we recover the action
~\eqn{actioncutoffsharp}.

A deformation of the product implies in general a deformation of the gauge group and,
more precisely, it determines a deformation of the whole enveloping algebra (Hopf
algebra). This is indeed what happens in our case. The Hopf algebra is effectively
deformed and it is shown to be inequivalent to the classical one for any non-trivial
choice of the cutoff function. This means that, although analytic cutoff functions  may
be reabsorbed with  a field redefinition from  the point of view of the algebra of
fields, the Hopft algebra symmetry cannot be mapped into the undeformed one.

The procedure illustrated can  be extended to nonabelian theories\footnote{To avoid
confusion we term ``noncommutative'' the theories on noncommutative
spacetimes, like the ones built using deformed products like the
Gr\"onevold-Moyal ones. By ``nonabelian'' we indicate the case in
which the gauge group is nonabelian, like ordinary Yang-Mills.
A theory can be nonabelian and noncommutative.} as well.

We also notice that a cutoff in momentum space is similar (but not exactly equivalent) to having a minimal length. Recently there has been interest in electrodynamics with minimal length, see for example~\cite{GaeteHelayelSpallucci} and references therein.

\section{{Momentum Cutoff via a Deformed Product}}
\setcounter{equation}{0}The renormalization group describes the
scaling of a field theory as a dimensional parameter is varied. The
parameter is responsible for the regularization of the  theory which would
otherwise have divergencies.  There are several
ways to implement the regularization (see for example~\cite[Chapt.~7.5]{Kaku} for
a quick comparison of the most popular schemes). The one used
in~\cite{Polchinski} is to consider a deformation of the free
propagator by a cutoff function in the momenta.

In~\cite{iraniani2} Ardalan, Arfaei, Ghasemkhani and Sadooghi proposed to implement the
cutoff with the use of a \emph{deformed commutative} product, employing the experience
gained in the study of noncommutative field theory. For simplicity let us examine for
the moment a scalar theory, the generalization to gauge theories will be done later in
Sect.~\ref{se:gauge}. The products considered (which we denote by $\star$) are
associative products among functions on spacetime with the property:
\be
\int \dd x f\star g = \int \dd x g\star f .\label{trace}
\ee
It is not actually crucial that the integral be done with the usual measure, but it is
necessary that there exist  an integral with the tracial property, i.e.\ invariant for
cyclic permutation of the factors. Another important property is the existence of
derivations, that is  operators satisfying the Leibnitz rule. For our case it is
sufficient to assume the Leibnitz rule for the usual derivations:
\be
\del_\mu (f\star g)=f\star\del_\mu g+ (\del_\mu f) \star g.
\ee
The most studied product of this kind is the Gr\"onewold-Moyal
product~\cite{Gronewold, Moyal}, which is noncommutative, and
reproduces the commutation rules of quantum mechanics adapted to
spacetime: $x^\mu\star x^\nu - x^\nu \star
x^\mu=\ii\theta^{\mu\nu}$.

\subsection{Poincar\'e Invariant star products}
Let us consider space-time symmetries and let us start imposing
translation invariance.  We will discuss Lorentz invariance as a
further restriction. In~\cite{GLV09} translation invariant
associative products were introduced and classified on the basis of
a suitable cohomology. The problem was further considered
in~\cite{iraniani,iraniani2}, and employed in the study of gauge
theories. Let us briefly review the subject.

For our purposes a generic star product is an associative product between functions on
$\real^d$ which depends on one or more parameters. In the limit in which these
parameters vanish the product becomes the usual pointwise product. We contemplate the
possibility that the star product be commutative. This will indeed be the main object of
interest.
\newcommand{\tran}{\mathcal T}
Let $(\mathcal{A},\mu)$ be the algebra of real functions with multiplication law
\be
\mu:\mathcal{A} \otimes \mathcal{A}\rightarrow \mathcal{A}.
\ee
 A translation invariant associative product may
be expressed as
\be
\mu(f\otimes h )(x)=(f\star h)(x)=\frac1{(2\pi)^{\frac d2}}\int\dd p^d \dd q^d \dd k^d
\e^{\ii p \cdot x} \tilde f(q)\tilde h(k) K(p,q,k) \label{intprod}
\ee
where $\tilde f(q)$ is the Fourier transform of $f$. $K$ can be a distribution and has
to implement translation invariance and associativity.  The product of $d$-vectors is
understood with the Minkowski or Euclidean metric: $p\cdot x=p_ix^i$. The usual
pointwise product is reproduced with the choice  $K(p,q,k)=\delta^d(k-p+q)$.
Defining the translation by a vector $a$ as $\tran_a(f)(x)=f(x+a)$,
by translation invariance of the product we mean the property
\be
\tran_a(f)\star \tran_a(h)=\tran_a(f\star h).
\ee
Performing the  Fourier transform we have
\be
\widetilde{\tran_a f}(q)=\e^{\ii a p}\tilde f(q)
\ee
so that translation invariance imposes on the  product~\eqn{intprod} the condition
\bea
&&\e^{\ii a\cdot p} \int \dd p^d \dd q^d \dd k^d \e^{\ii p \cdot x} \tilde f(q)\tilde
h(k)
K(p,q,k) =\nn\\
&=& \int \dd q^d \dd k^d \e^{\ii a\cdot q} \e^{\ii a\cdot k}\e^{\ii p \cdot x} \tilde
f(q)\tilde h(k) K(p,q,k).
\eea
At the distributional level this means
\be
K(p,q,k)=\e^{\ii(k-p+q)\cdot a}K(p,q,k)
\ee
and it is solved by
\be
K(p,q,k)=\mathcal{K}(p,q)\delta(k-p+q) \label{tinv}
\ee
with $\mathcal{K}$  a generic function to be further constrained.
In~\cite{GLV09} it was actually represented in exponential form, but
there is no particular reason for that, as pointed out
in~\cite{iraniani}. As we will see, a different representation
unveils a precious freedom in the choice of product representatives
in each equivalence class.

Strong  restrictions on ${K}$ come from the associativity requirement which reads
\be
\int\dd k^d K(p,k,q)K(k,r,s)=\int\dd k^d K(p,r,k)K(k,s,q). \label{cocyclecondition}
\ee
This is nothing but the usual cocycle condition in the Hochschild cohomology. See
\cite{GLV09} for details. Eq. \eqn{cocyclecondition} may be rephrased in terms of  a
condition for $\mathcal{K}$  \cite{iraniani}
\be
\mathcal{K}(p,q)\mathcal{K}(q,r)=\mathcal{K}(p,r)\mathcal{K}(p-r,q-r).\label{assocond}
\ee
Under rather mild assumptions of analicity  the most general solution is provided by
\cite{iraniani}
\be
\mathcal{K}(p,q)=H^{-1}(p) H(q) H(p-q) \e^{\ii  \alpha (p,q)}
\label{Kpq}
\ee
with $\alpha(p,q)$ a two cocycle in the appropriate cohomology \cite{GLV09} and $H(q)$
an arbitrary even real function. It can be shown that $\alpha(p,q)$  is constrained by
the associativity condition and the cyclicity  of the product \eqn{trace} to be of the
form
\be
\alpha(p,q)= \theta^{\mu\nu}p_\mu q_\nu +  \del\beta (p,q)=
\theta^{\mu\nu}p_\mu q_\nu+
\beta(q)-\beta(p)+\beta(p-q) \label{alphapq}
\ee
where we have emphasized the fact that it is defined up to a coboundary term, $\del
\beta(p,q)$, whose explicit form has been calculated. The function $\beta$ is a real odd function.

The matrix $\theta^{\mu \nu}$ is an antisymmetric and constant,
responsible for the noncommutativity of the product.  If $\theta=0$
the product is commutative. Commutative products are associated to
coboundaries.

The request of   Lorentz invariance further constrains the form of the product. Indeed
only the function $H$ survives, the term in $\theta$ being manifestly not invariant and
the function $\beta$ being an odd function of the modulus of momenta. Hence, the
requirement of full Poincar\'e invariance forces upon us the commutativity of the
product\footnote{From now on, with an abuse of notation, we will indicate by
$p,q,\ldots$ both the four-vector or its modulus. It will be clear from the context
which of the two we are meaning.}. We stress however that this does not mean that the
product is the usual pointwise one.

\subsection{Regularized product}

We will concentrate from now on the commutative, Poincar\'e  invariant case,  $\theta=0, \beta=0$.  We will refer to
field theories with the new product as {\it
commutative deformed field theories}, not to be confused with noncommutative ones. The
algebra of functions with the deformed product will be indicated with
$(\mathcal{A}_\star,\mu_\star)$ while the algebra of functions with point-wise product
will be denoted by $(\mathcal{A}_0,\mu_0)$.

The new star product, $\mu_\star$, acquires the form
\be
\mu_\star(f\otimes h)(x)=(f\star h )(x)=\int \frac{\dd^4
p}{(2\pi)^4}\frac{1}{H(p)} e^{\ii px} \int\frac{\dd^4 q}{(2\pi)^4}
[H(p-q)\tilde f(p-q)][H(q)\tilde h(q)] \label{starprod}
\ee
which is commutative and Poincar\'e invariant but non-local. $H=1$ corresponds to the
ordinary point-wise product $\mu_0$. It is possible to show that the deformed product
enjoys the property
\be
\int \dd^4 x (f\star h)(x)=\int \frac{\dd^4 p}{(2\pi)^4} H^2(p)
\tilde f(-p) \tilde h(p) \label{integral}
\ee
where we have used  $H(0)=1$.  This is a consequence of the request that the algebra of functions be unital with
ordinary unity
\be
f\star 1= 1\star f= f.
\ee
The identity \eqn{integral} will be useful in computing the deformed QED action in momentum space  and will play a
crucial role in the limiting process that
we will need to properly define the cutoff action~\eqn{actioncutoffsharp}.

It was argued  in~\cite{iraniani2} that the freedom acquired
with the product deformation through the function $H(p)$, may be
exploited to regularize quantum field theories. In that proposal the
function $H$ plays the role of a multiplicative cut off function in
momentum space. Let us analyze better this. The presence of $H^{-1}$
in~\eqn{starprod} implies the fact that the cutoff function cannot
vanish, except possibly on a measure zero set if one considers it
as a distribution. In particular it is impossible to consider a
function which vanishes identically outside a sphere of finite
radius. This is ultimately a consequence of the nonlocal nature of
the convolution product of the Fourier transform. The idea suggested
in~\cite{iraniani2} to consider only functions whose Fourier
transform vanishes outside a sphere of radius $\Lambda'\gg\Lambda$
does not really work since the convolution product is nonlocal, and
even if the starting functions have this property their product will
not have it. It is easy to see that, since $\Theta_\Lambda(p)$ is
not translationally invariant there is no way to obtain an
associative product.

On the other hand, as long as the cut-off
function is analytic the new product belongs to the same cohomology
class as the point-wise product.  In that case   it was shown
in~\cite{GLV08,GLV09} that the ultraviolet
behaviour of the deformed quantum field theory remains unmodified
with respect to ordinary quantum field theory. In other terms, a non
vanishing cut-off function provides essentially a field
redefinition, basically the multiplication of the Fourier transforms
of the fields by the function $H$, which cannot improve  the
ultraviolet regime.
Indeed we have, for each invertible  $H$, an isomorphism
between the deformed and undeformed algebra of fields,
\be
\varphi: (\mathcal{A}_\star, \mu_\star)\rightarrow (\mathcal{A}_0,
\mu_0)
\ee
which, in momentum space, reads
\beqa
\widetilde{\varphi(f)}(p)&=&H(p) \tilde f(p) \label{isomorph}\\
\widetilde{\varphi(f\star g)}(p)&=&\widetilde{\varphi(f)}\bullet
\widetilde{\varphi(g)}
\eeqa
with $\bullet$ the  undeformed convolution. {\footnote{ However we will show in
Sect.~\ref{se:inequiv} that this equivalence does not extend to the
algebra of symmetries.  }}

{Despite these warnings we will  show in Sect.~\ref{se:limit} that
the regularized theory with the sharp cutoff can be properly defined
as the limit $H(p)\rightarrow \Theta_\Lambda(p)$ of well defined
theories with analytic  cutoff.}

As noted already in~\cite{iraniani2} the theory with a sharp  cutoff
is akin to putting the theory on a lattice, although in that case
the momentum is periodic, which is not true in this case. A closer
analogy is with quantum field theories on fuzzy spaces (see
\cite{fuzzy} and refs therein) where the regularization is achieved
without destroying the symmetry. The realization of fuzzy  noncompact spaces may be found in \cite{discofuzzy} with an
application to the regularization of scalar field theory.  A related case is the implementation of the
cutoff via the eigenvalues of a generalized Dirac operator.  This is
the basis of the spectral action~\cite{SpectralAction} and of the
finite mode regularization (see for example~\cite{AndrianovBonora,
Fujikawa, AndrianovKurkovLizzi}).

\section{Deformed gauge symmetry \label{se:gauge}}
\setcounter{equation}{0} It is further proved in the
article~\cite{iraniani2} that the  deformation procedure can be implemented for
gauge theories and that the gauge symmetry is preserved by the
deformation adopted, although in a modified form. What remained to
be understood is the nature of the symmetry.

 Let $f,h\in
(\mathcal{A}_0,\mu_0)$. The pointwise product may be defined in
terms of the convolution product in momentum space
\be
(\tilde f \bullet \tilde h)(p)=\int \frac{\dd^4 q}{(2\pi)^4} \tilde
f(p-q)\tilde h(q)\label{conv}
\ee
so that
\be
\mu_0(f\otimes h)(x)=(f\cdot h) (x) = \int \frac{\dd^4
p}{(2\pi)^4}\e^{\ii px} (\tilde f \bullet \tilde h)(p).
\ee
Analogously we introduce the deformed convolution product
$\bullet_*$
\be
(\tilde f \bullet_\star \tilde h)(p)= \frac{1}{H(p)} \int
\frac{\dd^4 q}{(2\pi)^4} H(p-q)\tilde f(p-q)\tilde H(q)
h(q)\label{defoconv}
\ee
so to have
\be
\mu_\star(f\otimes h)(x)=(f\star h )(x) = \int \frac{\dd^4
p}{(2\pi)^4}\e^{\ii px} (\tilde f \bullet_\star \tilde h)(p).
\label{defoprod}
\ee

Ordinary  gauge theories  with gauge group\footnote{In our notation $G$ is the Lie
group, with Lie algebra $\mathfrak{g}$. The hatted objects $\widehat G,\;
\widehat{\mathfrak{g}}$ indicate respectively the group and the algebra of gauge
transformations, i.e.\ functions from spacetime in the group or algebra.} $\widehat G$
are modified replacing the point-wise product with the non-local product~\eqn{defoprod}.
The resulting field theories are invariant under the deformed gauge transformations
\be
\phi(x) \longrightarrow g_\star(x)\triangleright_\star \phi(x)=\exp_\star\left(i
\alpha^i (x) T_i\right) \triangleright_\star \phi (x). \label{gaugetransf}
\ee
 $\triangleright_\star$ indicates generically the action of the group and  later of the   algebra,
which implies (for the nonabelian case) both a matrix multiplication and a $\star$
product. In the U(1)  case the notation is  redundant, it being $(\alpha
\triangleright_\star\phi)(x)=\alpha(x)\star\phi(x)$, but the more general notation will
be useful later on. Here $T_i$ are the Lie algebra generators, $T_i \in \mathfrak{g}$
and the gauge group elements $g_\star(x)$ are defined as star exponentials
\be
g_\star(x)=\exp_\star\left(i \alpha (x)^i T_i\right)= 1+ i \alpha^i
(x) T_i - \frac{1}{2}(\alpha^i\star \alpha^j)(x) T_i T_j + \ldots
\label{starexp}
\ee
At the infinitesimal level we have then
\be
\phi(x) \longrightarrow  \phi(x) +i (\alpha\triangleright_\star \phi)(x)
\label{infgaugetransf}
\ee
with
\be
(\alpha\triangleright_\star\phi)(x)=i \left(\alpha^j(x)\star (T_j\triangleright \phi)
\right)(x)
\ee
and $T_j$ is in the appropriate representation to  the field $\phi$.
The construction is valid for fields with non zero spin as well. In
\cite{iraniani2} quantum electrodynamics is explicitly considered.

The deformed Lie multiplication reads
\be
[\alpha,\tilde\alpha]_\star(x)=(\alpha \star \tilde \alpha) (x) - (\tilde
\alpha\star\alpha) (x). \label{defolie}
\ee

\noindent{\bf  Remark}\ In noncommutative field theory the
definition~\eqn{gaugetransf} is problematic because we have
\bea
(\alpha \star \tilde \alpha) (x) - (\tilde \alpha\star\alpha) (x)&=&
\left((\alpha^i \star \tilde\alpha^j )(x)+(\tilde \alpha^j\star
\alpha^i)(x)\right) [ T_i, T_j]\nonumber\\&&+\left ((\alpha^i\star
\tilde\alpha^j) (x)-(\tilde \alpha^j\star \alpha^i)(x)\right)
\{T_i,T_j\}
\eea
which only closes for the group $U(N)$ in the adjoint and
fundamental representations. This problem is solved for example in
twisted gauge theories(see for example~\cite{WSS}) where the star
product is induced by a twist operator and the gauge transformations
are twist-deformed. In the present case the
definition~\eqn{gaugetransf} is perfectly viable for any Lie group,
because the product is commutative, therefore the term proportional
to the anticommutator $\{T_i,T_j\}$ vanishes.

\subsection{The Hopf algebra structures of ordinary gauge theory}
The action of the group on the fields, when products of fields are considered, involves
not only the Lie-algebra structure, but also the full Hopf-algebra structure, although
for the pointwise multiplication the latter structure is trivial. This is not the place
to describe all of aspects of the theory of Hopf algebras (for an introduction see for
example~\cite{chari}). In the following we will just recall the essential definitions.

When acting on the product of fields with a gauge transformation we need to extend the
action of the gauge group. This is  obtained via the  coproduct.
Since we will mostly deal with infinitesimal gauge transformations, $\alpha(x)\in
\widehat{\mathfrak{g}}$, we review here the coproduct for the infinitesimal gauge generators while the Hopf algebra
structure of finite gauge transformations and its deformation are discussed  in the Appendix  \ref{appendix}.

The coproduct  is properly defined on the universal enveloping algebra of  $ \widehat
{\mathfrak{g}}$,
\be
\Delta:U( \widehat {\mathfrak{g}})\otimes U( \widehat {\mathfrak{g}})\longrightarrow  U(
\widehat {\mathfrak{g}}).
\ee
Its explicit form  may be obtained on asking that the action of the group be
an automorphism of the algebra $\mathcal{A}$, i.e. that it be compatible with the
multiplication law in $\mathcal{A}$. We have
\be
\alpha\triangleright \mu\circ(f\otimes h) =
\mu\circ(\rho\otimes\rho)(\Delta(\alpha))\circ(f\otimes h).
\ee
 From the request that the coproduct  be compatible with the ordinary point-wise product
we obtain,
\beqa \label{compat_algebra}
\Delta_0 (\alpha) (f\otimes h) &=& (\alpha\otimes \mathrm{id} \, + \mathrm{id}\,\otimes
\alpha)(f\otimes h) = \alpha \triangleright f\otimes h + f \otimes \alpha \triangleright
h\\
\Delta_0 (\mathrm{id}_{U(\widehat{\mathfrak{g}})})(f\otimes
h)&=&(\mathrm{id}_{U(\widehat{\mathfrak{g}})}\otimes
\mathrm{id}_{U(\widehat{\mathfrak{g}})})(f\otimes h).
\eeqa
This endows $U(\widehat{\mathfrak{g}})$ with a cocommutative Hopf algebra structure,
provided we define  an algebra homomorphism
$\epsilon$ called the counit and an algebra antihomomorphism $S$, the antipode,
\beqa
       \epsilon : U(\widehat{\mathfrak {g}}) &\rightarrow& \mathbb{C}
\;\;\;\;\;\;\;\;\;\;
\epsilon(\alpha) = 0 \;\;\;\;\forall \alpha\in \widehat{\mathfrak {g}} \\
         S : U(\widehat{\mathfrak {g}})&\rightarrow & U(\widehat{\mathfrak {g}})\;\;\;
S(\alpha) = -\alpha \;\;\forall \alpha\in \widehat{\mathfrak {g}}.
\eeqa
It is easily verified that $\Delta_0, S, \epsilon,$ so defined are mutually compatible, that is they satisfy the
conditions
\eqn{cond1}-\eqn{cond3}.
 From \eqn{compat_algebra} we have
\be
\alpha \triangleright (f\cdot h)(x)= \mu_0 \circ \Delta_0 (\alpha)(f\otimes h)(x)=\left
( (\alpha\triangleright f ) \cdot h\right)(x)+ \left ( f \cdot (\alpha\triangleright h)
\right)(x). \label{leibnitzrule}
\ee
This is  the usual Leibnitz rule, with
\be
(\alpha\triangleright f )(x)=\left(\alpha^i \cdot (T_i\triangleright f)\right)(x) .
\ee

\subsection{The deformed Hopf algebra}
When the point-wise multiplication is deformed, the Leibnitz rule which is encoded in
the coproduct  changes accordingly. On generalizing \eqn{leibnitzrule} we have
\be
\alpha \triangleright (f\star h)(x)= \mu_\star \circ \Delta_\star (\alpha)(f\otimes
h)(x) \label{defolieb}
\ee
with a deformed coproduct to be determined. From the expression of the deformed product
\eqn{defoprod} and the deformed gauge action \eqn{gaugetransf} we read off the
expression of the new coproduct
\be
\Delta_\star(\alpha)(f\otimes g)= \left({\alpha}_\star\otimes
\mathrm{id}+\mathrm{id}\otimes {\alpha}_\star\right)(f\otimes
g)=\alpha\triangleright_\star
 f\otimes g+ f\otimes\alpha \triangleright_\star g \label{coprodstar}
\ee
and
\be
(\alpha\triangleright_\star  f)(x)=(\alpha^i\star T_i\triangleright f)(x)
\ee
where the action of the ungauged  Lie algebra in the appropriate representation is not
modified. It can be verified that our definition is consistent  with the coassociativity
condition \eqn{cond1}. Moreover, the coproduct $\Delta_\star$ is cocommutative according
to the definition in \eqn{coco}. We then define the antipode and the counit
\beqa
\left(S_\star(\alpha)\triangleright f\right)(x)&=&-(\alpha\triangleright_\star f)(x)
\label{antipode*}\\
\epsilon_\star(\alpha)&=&\epsilon(\alpha)=0. \label{counitstar}
\eeqa
 These definitions are consistent with Eqs. \eqn{cond2},
\eqn{cond3}.

To summarize, the deformed universal enveloping algebra
$U_\star(\widehat{\mathfrak{g}})$ is a cocommutative Hopf algebra with deformed Lie
multiplication, $[\cdot, \cdot]_\star$, deformed coproduct $\Delta_\star$, deformed
antipode, $S_\star$, and undeformed counit $\epsilon$.

\section{Deformed field theories}
\setcounter{equation}{0}

\subsection{The deformed Hopf symmetry}

It is not difficult  to verify that the action of quantum electrodynamics equipped with
the star product \eqn{starprod}
\be
S_{H}= \int \dd^4 x \left\{-\bar\psi \star (i\gamma^\mu\del_\mu
-m)\psi +e \bar\psi\star\gamma^\mu
A_\mu\star\psi+\frac{1}{4}F_{\mu\nu}\star F^{\mu\nu}
+\frac{1}{2\xi}(\del_\mu A^\mu)^{\star 2}\right\}
\ee
is invariant under the deformed {U(1) Hopf algebra, when  the deformed coproduct
\eqn{coprodstar} is used. On generalizing  \eqn{gaugetransf} the deformed gauge
transformations for matter and gauge fields explicitly read
\be
\psi(x) \rightarrow \psi(x)+ ie(\alpha \star
\psi)(x),\;\;\;\bar\psi(x) \rightarrow \bar\psi(x)- ie(\alpha \star
\bar\psi)(x),\;\;\; A_\mu(x)\rightarrow A_\mu(x)-\del_\mu \alpha(x).
\ee
They become in momentum space
\beqa \delta \tilde\psi(p)&=&\ii e H^{-1}(p) \int
\frac{\dd^4q}{(2\pi)^4} H(p-q)\tilde \alpha(p-q)
H(q)\tilde \psi(q)\\
\delta \tilde{\bar{\psi}}(p)&=&-\ii e H^{-1}(p) \int
\frac{\dd^4q}{(2\pi)^4} H(p-q)\tilde \alpha(p-q)
H(q)\tilde{\bar{\psi}}(q)\\
\delta \tilde A_\mu(p)&=&\ii p_\mu \tilde\alpha(p).
\eeqa
On using \eqn{integral} the action is rewritten as
\beqa
S_{H}&=& \int \frac{\dd^4 p}{(2\pi)^4}\; {\Biggl\{ } H(p)
\tilde{\bar{\psi}}(-p)(\gamma_\mu p^\mu+m)H(p)\tilde \psi(p)\biggr.\nn\\
 &&+ e  H(p)
\tilde{\bar{\psi}}(-p)\int \frac{\dd^4q}{(2\pi)^4} H(p-q)
\gamma^\mu\tilde A_\mu(p-q)
H(q)\tilde \psi(q) \nn\\
 &&+  \frac{1}{4}H^2(p) \left(\tilde F_{\mu\nu}(-p)\tilde
F^{\mu\nu}(p)+\frac{1}{2\xi} p_\mu \tilde A^\mu(-p)\, p_\nu \tilde A^\nu(p) \right)
\Biggr\}. \label{action}
\eeqa
Let us notice that, although the modified product~\eqn{starprod}
contains the inverse function of $H$, this has disappeared from the
action thanks to the identity \eqn{integral}.

 Therefore, if $H$ can be chosen to be   a cutoff function, the theory is fully
regularized. It is gauge invariant by construction (with respect to the deformed symmetry
discussed above), but it is also possible to prove the relevant Ward identities, the
derivation mirrors the usual one and is described in detail in~\cite[App.~B]{iraniani2}.

\subsection{Limit of Hopf Algebras and non-analytic cutoff}\label{se:limit}

For the product~\eqn{starprod} to be defined it is necessary that
the function $H(p)$ do not vanish anywhere. But in that case, as we
have already argued, the deformation is not very interesting because
the new product is isomorphic to the point-wise one.  The
action~\eqn{action} however can be defined for arbitrary cutoff
functions, including those which identically vanish for $p^2$ larger
than some scale. In this case the theory cannot be obtained from the
ordinary theory via a field redefinition. For example in the event
of {non-analytic}  cutoff there is an actual loss of information,
all momenta above $\Lambda$ do not contribute.  This is the
interesting case, but we cannot simply cut the momenta, since we
would lose associativity of the product. We thus consider a sequence
of analytic cutoff functions  which converge to the sharp cutoff
$\Theta_\Lambda(p)$ .
\be
H_\epsilon(p)\to\Theta_\Lambda(p) \label{htheta}.
\ee
A possible choice is for example  the following sequence of functions
\be
H_\epsilon(p)=\frac12-\frac12\tanh\left(\frac{p^2-\Lambda^2}{\epsilon\Lambda^2}\right).
\ee
They converge to $\Theta_\Lambda(p)$ in the limit $\epsilon\to 0$.
At each stage of the limiting procedure the action \eqn{action} preserves the symmetries, while converging to the
cutoff-action \eqn{actioncutoffsharp} introduced at the beginning of this article, in the limit \eqn{htheta}.

The theory with the sharp cutoff cannot be defined with a deformed
product, nevertheless, being a limit of Hopf-gauge invariant
theories, it enjoys their symmetries, and the proof of the Ward
identities of~\cite{iraniani2} still goes through.

The limits we are taking here are  to be understood in the weak (nonuniform) sense.  At
any stage the theory satisfies the Ward identities, and it has the full Hopf invariance.

\subsection{Inequivalence of $U(\hat{\mathfrak{g}})$ and
$U_\star(\hat{\mathfrak{g}})$}\label{se:inequiv}

The deformed symmetry of the QED action~\eqn{action}, the Hopf algebra
$U_\star(\widehat{\mathfrak{u}(1)})$ is  commutative (because $U(1)$ is Abelian) and
cocommutative, because the deformed coproduct satisfies \eqn{coco},
 exactly as $U(\widehat{\mathfrak{u}(1)})$. Therefore we can ask whether the two Hopf algebras be
or not equivalent. A precise definition of equivalence is in terms of homomorphisms. Two
Hopf algebras $(\mathcal{A},m,\Delta,S)\;(\mathcal{B},m',\Delta',S')$ are equivalent if
there exists a map
$$\varphi: \mathcal{A}\longrightarrow\mathcal{B}$$
which is
\begin{enumerate}
\item an algebra homomorphism,
$$\varphi\circ m=m'\circ\varphi $$
\item a coalgebra homomorphism,
$$(\varphi\otimes\varphi)\circ \Delta=\Delta'\circ \varphi$$
\item a Hopf algebra homomorphism, that is
$$\varphi\circ S=S'\circ \varphi. $$
\end{enumerate}
Moreover it has to be compatible with the action on the algebra of fields. That is
\be
\widetilde{\varphi(\alpha\triangleright_\star \phi)}(p)=\widetilde{\varphi(\alpha)}\triangleright \widetilde{\varphi(\phi)} \label{comp}
\ee
It is not difficult to see that in the case of analytical $H$  there exists a
map between $U(\mathfrak{g})$ and $U_\star(\mathfrak{g})$ which is an algebra
homomorphism. Eq. \eqn{comp} actually imposes that it be  the map already defined for the algebra of fields \eqn{isomorph}.
This is the field redefinition we have already alluded. It is however not a Hopf algebra
morphism because it is not difficult to check  that it does not satisfy properties 2 and 3 above.

We conclude that although the deformed algebra with nonlocal product
is isomorphic to the standard one with point-wise product, the
deformed symmetry of the action~\eqn{action} is a genuine new
symmetry.

\section{Conclusions}
\setcounter{equation}{0}

Techniques borrowed from field theories on noncommutative spaces have been used in the
context of the regularization of ordinary field theory. The main tool has been a
commutative, Poincar\'e invariant deformed product. The deformed theory has been seen to
possess a deformed  symmetry, not only in the Lie algebra structure, but at the full
Hopf algebra level. This regularized electrodynamics is an instance of  the simplest
possible Hopf algebra invariance, commutative, cocommutative, associative and
coassociative. This is the ``hidden'' symmetry alluded to in~\cite{iraniani2}. The
resulting theory is fully gauge invariant, and the general technique makes no
distinction between Abelian and nonabelian symmetries.

The regularization depends on a cutoff function which, even if necessarily non-vanishing
for the definition of the product, may be taken to converge to a sharp cutoff. In this
limit the regularization is not just a field redefinition. The effective limiting theory
does indeed depend only on the sector of the theory below the cutoff.  At each stage the
theory possesses internal and spacetime symmetries.

The deformation of products seems a promising tool, not only for renormalization issues,
but also for the performance of actual calculations and may turn out to be particularly
useful for nonabelian theories.}


\paragraph{Acknowledgments}
We thank Francesco D'Andrea for discussions. F.~Lizzi acknowledges
support by CUR Generalitat de Catalunya under project FPA2010-20807
and the {\sl Faro} project {\sl Algebre di Hopf, differenziali e di
vertice in geometria, topologia e teorie di campo classiche e
quantistiche} of the Universit\`a di Napoli {\sl Federico II}.
P.~Vitale acknowledges a grant from the European Science Foundation
under the research networking project "Quantum Geometry and Quantum
Gravity", and partial support by GDRE GREFI GENCO.

\setcounter{section}{0}
\renewcommand{\thesection}{\Alph{section}}
\setcounter{equation}{0}

\section{\label{appendix}Appendix}
We review in this appendix the Hopf algebra structure of finite (as opposed to infinitesimal) gauge transformations, and
introduce  their deformation.

Let $\mathbb{C}\widehat
G$ be the group algebra of $\widehat G$ over the complexes, with multiplication $m$. The
coproduct is a homomorphism from $\mathbb{C}\widehat G$ to $\mathbb{C}\widehat G\otimes
\mathbb{C}\widehat G$
\be
\Delta:\mathbb{C}\widehat G \rightarrow \mathbb{C}\widehat G\otimes \mathbb{C}\widehat
G, \;\;\; \forall g \in \widehat G, f,h\in \mathcal{A},\;\; g\triangleright (f\otimes h)
= (\rho\otimes\rho)(\Delta(g))\circ(f\otimes h)
\ee
with $\rho$ a representation of $\mathbb{C}\widehat G$ on the algebra of fields.
 $\triangleright$ indicates the action of the gauge group. The
explicit form of the coproducts is  obtained on asking that the action of the group be
an automorphism of the algebra $\mathcal{A}$, i.e. that it be compatible with the
multiplication law in $\mathcal{A}$,
\be
g \triangleright \mu (f\otimes h) (x)= \mu(\Delta(g)\triangleright (f\otimes h)(x).
\label{compatibility}
\ee
The ordinary coproduct, compatible with the point-wise
multiplication, is easily obtained by Eq.~\eqn{compatibility} to be,
for  group-like elements
\be
\Delta_0(g)=g\otimes g \label{delta0}.
\ee
The rule is compatible with the multiplication $m$, therefore it can
be   uniquely extended to the whole of $\mathbb{C}\widehat G$. The
coproduct $\Delta_0$ is cocommutative, i.e. it satisfies
\be
\tau\circ\Delta_0=\Delta_0 \label{coco}
\ee
with $\tau$ the permutation operator, $\tau (f\otimes h)= h\otimes f$.

The group algebra $(\mathbb{C}\widehat G,m,\Delta)$ is a bialgebra. It is turned into a
Hopf algebra if we also have an algebra homomorphism, $\epsilon$ called the counit
\be
\epsilon :\mathbb{C}\widehat G \rightarrow \mathbb{C} \;     \epsilon(g) = 1 \;\;\forall
g \in \widehat G
\ee
and an algebra antihomomorphism $S$, the antipode
\be
S : \mathbb{C}\widehat G\rightarrow\mathbb{C}\widehat G\;\; S(g) = g^{-1}  \;\; \forall
g\in \widehat G.
\ee
Both can be uniquely extended to the whole group algebra.

For a given  Hopf algebra $\mathcal{H}$ the coproduct, the counit and the antipode have
to satisfy the relations below:
\beqa
(\Delta\otimes \mathbb{I})\circ \Delta &=& (\mathbb{I}\otimes\Delta) \circ \Delta \label{cond1}\\
(\mathbb{I}\otimes\epsilon)\circ \Delta&=&\mathbb{I} (\epsilon\otimes \mathbb{I})\circ
\Delta \label{cond2}\\
 m\circ (\mathbb{I}\otimes S)\circ \Delta&=& m\circ(S\otimes \mathbb{I})\circ
\Delta=1_\mathcal{H} \circ\epsilon .\label{cond3}
\eeqa
Eq.~\eqn{cond1} is the coassociativity condition. These relations have to be
verified when we  introduce the Hopf algebra structures of the deformed symmetry.

To conclude this section  we briefly describe the Hopf algebra structure of deformed gauge
transformations.

In total analogy with the universal enveloping algebra,  we can deform the group
algebra, $\mathbb{C}\widehat{G}$, together with its structures and co-structures. Group
elements $g(x)$ are replaced by star exponentials as in Eq. \eqn{starexp}.  The group
multiplication $m$ is thus replaced by $m_\star$
\be
m_\star\circ (g_\star\otimes \tilde g_\star) =  g_\star\star \tilde g_\star.
\ee
The deformed coproduct is obtained as previously, on requesting that it be an
automorphism for the algebra of fields $(\mathcal{A}_\star, \mu_\star)$. The antipode is
obtained by consistency with the coproduct. They read  respectively
\beqa
\Delta_\star(g) (f\otimes h)&=& g_\star \triangleright_\star f\otimes h + f\otimes
g_\star\triangleright_\star h \\
S_\star (g) \triangleright_\star f&=& (g_\star^{-1})\triangleright_\star f
\eeqa
and the counit is undeformed. It can be verified that these
definitions satisfy the consistency conditions~\eqn{cond1}
and~\eqn{cond3}.

\end{document}